\documentclass{article}

\usepackage[numbers]{natbib}
\usepackage{rotating}
\usepackage{graphicx}

\newcommand{\bin}[2]{\left(\begin{array}{c} \!\!#1\!\! \\  \!\!#2\!\! \end{array}\right)}
\newcommand{\troisj}[3]{\left(\begin{array}{ccc}#1 & #2 & #3 \\ 0 & 0 & 0 \end{array}\right)}
\newcommand{\sixj}[6]{\left\{\begin{array}{ccc}#1 & #2 & #3 \\ #4 & #5 & #6 \end{array}\right\}}
\newcommand{\neufj}[9]{\left\{\begin{array}{ccc}#1 & #2 & #3 \\ #4 & #5 & #6 \\ #7 & #8 & #9 \end{array}\right\}}
\newcommand{\ff}[3]{\mathcal{F}^{#1}(#2,#3)}
\newcommand{\gd}[3]{\mathcal{G}^{#1}(#2,#3)}

\begin{document}

\huge

\vspace{3cm}

\begin{center}
Statistical properties of levels and lines in complex spectra
\end{center}

\vspace{0.5cm}

\Large

\begin{center}
A tribute to Jacques Bauche and Claire Bauche-Arnoult
\end{center}

\vspace{0.5cm}

\large

\begin{center}
Jean-Christophe Pain\footnote{jean-christophe.pain@cea.fr (corresponding author)} and Franck Gilleron
\end{center}

\vspace{0.2cm}

\normalsize

\begin{center}
CEA, DAM, DIF, F-91297 Arpajon, France
\end{center}

\vspace{0.5cm}

%===============================================================================
%		ABSTRACT
%===============================================================================

\begin{center}
{\bf Abstract}
\end{center}

We review recent developments of the statistical properties of complex atomic spectra, based on the pioneering work of Claire Bauche-Arnoult and Jacques Bauche. We discuss several improvements of the statistical methods (UTA, SOSA) for the modeling of the lines in a transition array: impact of high-order moments, choice of the distribution (Generalized Gaussian, Normal Inverse Gaussian) and corrections at low temperatures. The second part of the paper concerns general properties of transition arrays, such as propensity rule and generalized $J$-file sum rule (for E1 or E2 lines), emphasizing the particular role of the $G^1$ exchange Slater integral. The statistical modeling introduced by J. Bauche and C. Bauche-Arnoult for the distribution of the $M$ values (projection of total angular momentum $J$) in an electron configuration, written $P(M)$, was extended in order to account for configurations with a high-$\ell$ spectator and a new analytical formula for the evaluation of the number of E1 lines with a wider range of applicability was derived.

\section{INTRODUCTION}

A method for calculating the quantum bound states of an ion consists in solving Schr\"odinger or Dirac equation for the one-electron states in an effective central potential, building the $N$-electron states as linear combinations of Slater determinants and obtaining the coefficients of the development by diagonalization of the hamiltonian \cite{Cowan81}. Such a detailed method (so-called Slater-Condon) has often a high numerical cost due to the multiplicity of accessible states. For example, in the  ion W$^{30+}$, the main complex $(n=1)^2 (n=2)^8 (n=3)^{18} (n=4)^{16}$ contains 601 080 390 quantum states ($\alpha JM$) and 49 309 974 degenerate levels ($\alpha J$). The statistical methods proceed differently \cite{Moszkowski62,Layzer63}; they consist in averaging the perturbative hamiltonian on well-chosen quantum numbers. No diagonalization is performed and the invariance of the trace of the hamiltonian enables one to obtain compact formulas for average energies. The idea of C. Bauche-Arnoult, J. Bauche and M. Klapisch \cite{Bauche-Arnoult79-82-85} was to extend those methods to more complicated operators to calculate, for instance, the moments of the energies of a group of lines weighted by transition strengths. When physical broadening mechanisms (Doppler, Stark, ...) are important, the lines merge together. Moreover, explicit quantum calculations can be inappropriate, e.g. if the hamiltonian or its eigenvalues are not known with a sufficient precision, and global methods may reveal physical properties hidden by a detailed treatment of levels and lines (``the tree can hide the forest'') \cite{Bauche90}. The precision and computation times of statistical methods are adjustable, from configurations to superconfigurations and the calculations are robust, making possible the generation of opacity tables (parametrized by $Z$, $T_e$, $N_e$). Global methods such as UTA (Unresolved Transition Arrays) and SOSA (Spin-Orbit Split Arrays) are well suited for complex configurations (high $Z$) in local-thermodynamic-equilibrium (LTE) or non-LTE plasmas, but they are not always precise enough for accurate Rosseland means or to interpret high-resolution spectra. We tried (modestly) to extend their validity range and to find some regularities \cite{Pain13} or to use general properties such as the propensity law in order to develop heuristic approximate models \cite{Gilleron07}. We paid a particular attention to the role of the $G^1$ exchange Slater integral, and the usefulness of emissive and receptive zones of configurations \cite{Bauche83}. We also developed new methods for estimating, exactly or approximately, the number of lines in E1 and E2 transition arrays.
                                                                                                                                                                                                                                                                  
\section{IMPROVEMENT OF STATISTICAL MODELS}

\subsection{Beyond the Gaussian assumption}

\begin{figure}[h]
\begin{center}
\includegraphics[width=210pt]{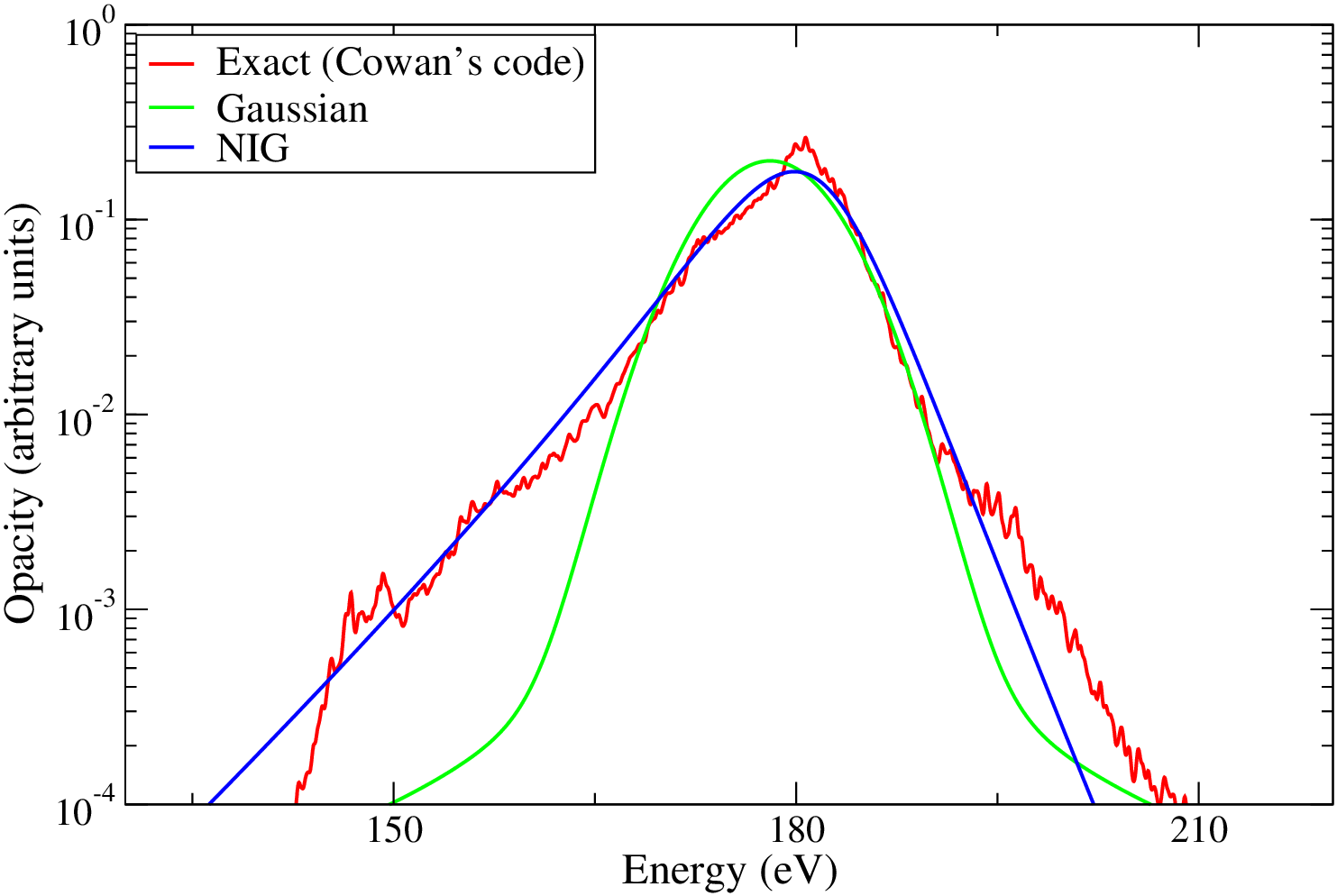}
  \caption{Different modelings of transition array $3p^53d^5\rightarrow 3p^53d^44p$ in Ge XII.}\label{wingsa}
\end{center}
\end{figure}

\begin{figure}[h]
\begin{center}
\includegraphics[width=190pt]{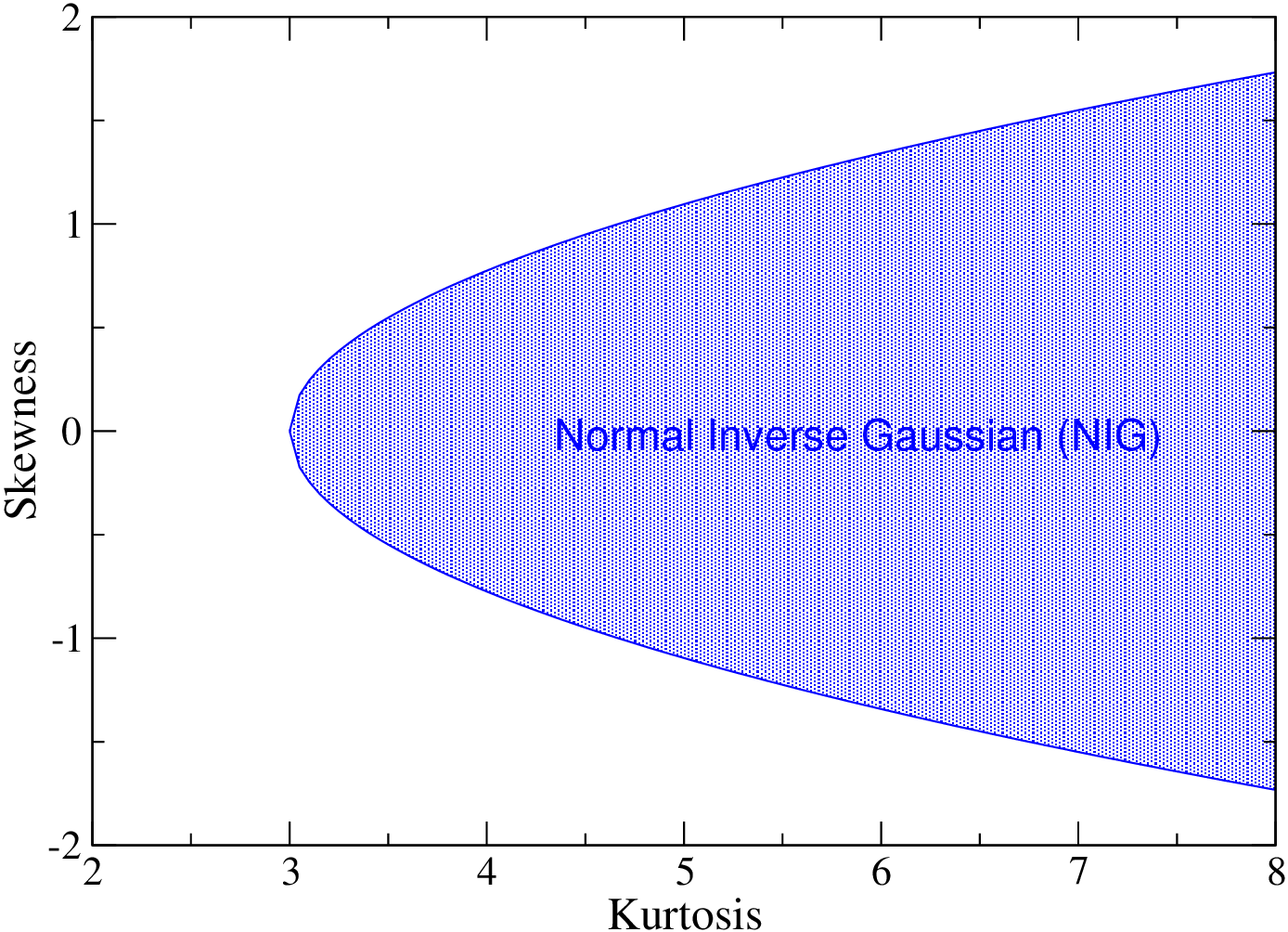}
  \caption{Validity domain of the NIG function in the $\left(\alpha_3,\alpha_4\right)$ representation (right).}\label{wingsb}
\end{center}
\end{figure}

\begin{figure}[h]
\begin{center}
\includegraphics[width=190pt]{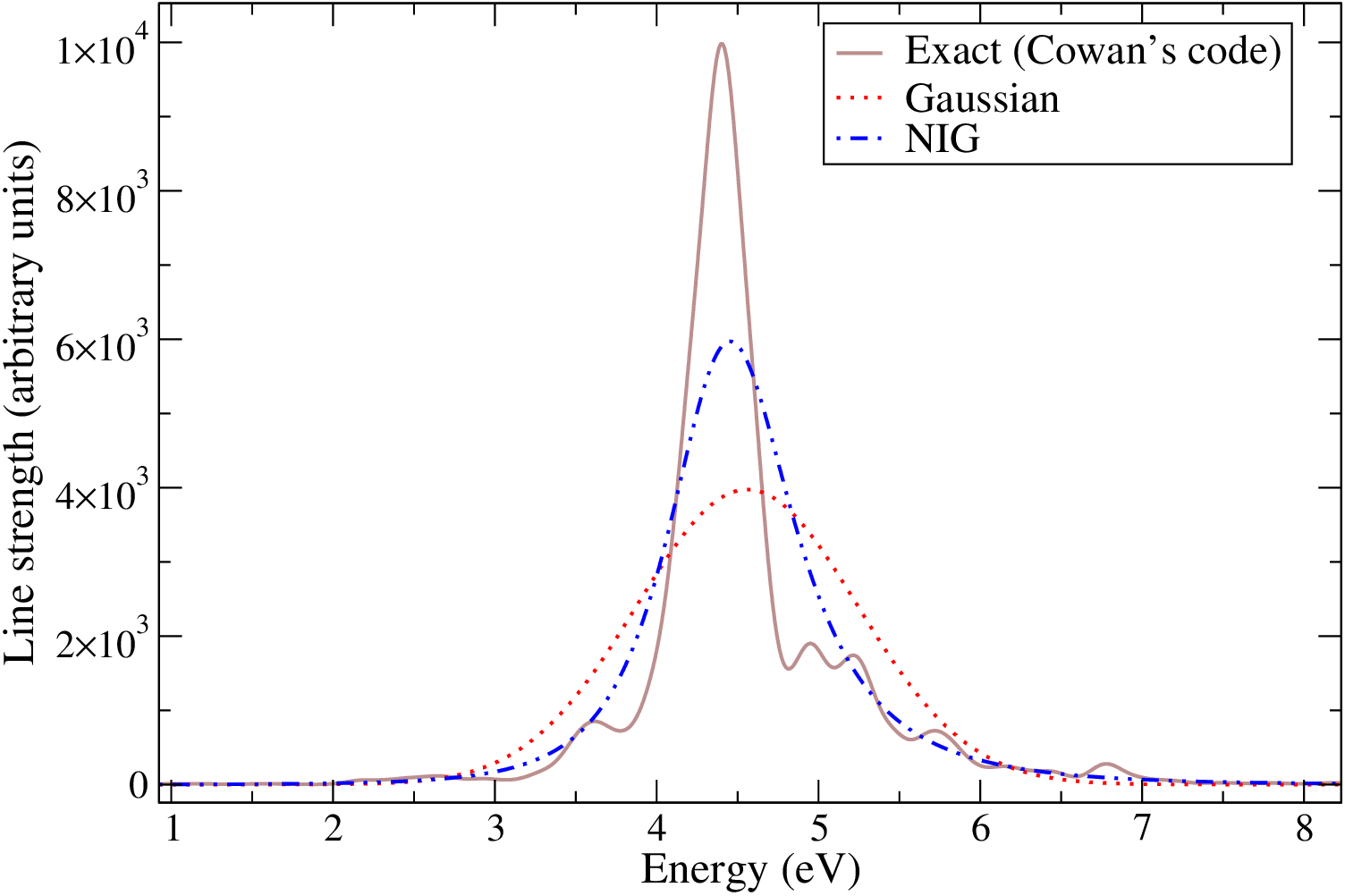}
  \caption{Different modelings of transition arrays $3d^34p\rightarrow 3d^34d$ in V II.}\label{ggniga}
\end{center}
\end{figure}

\begin{figure}[h]
\begin{center}
\includegraphics[width=195pt]{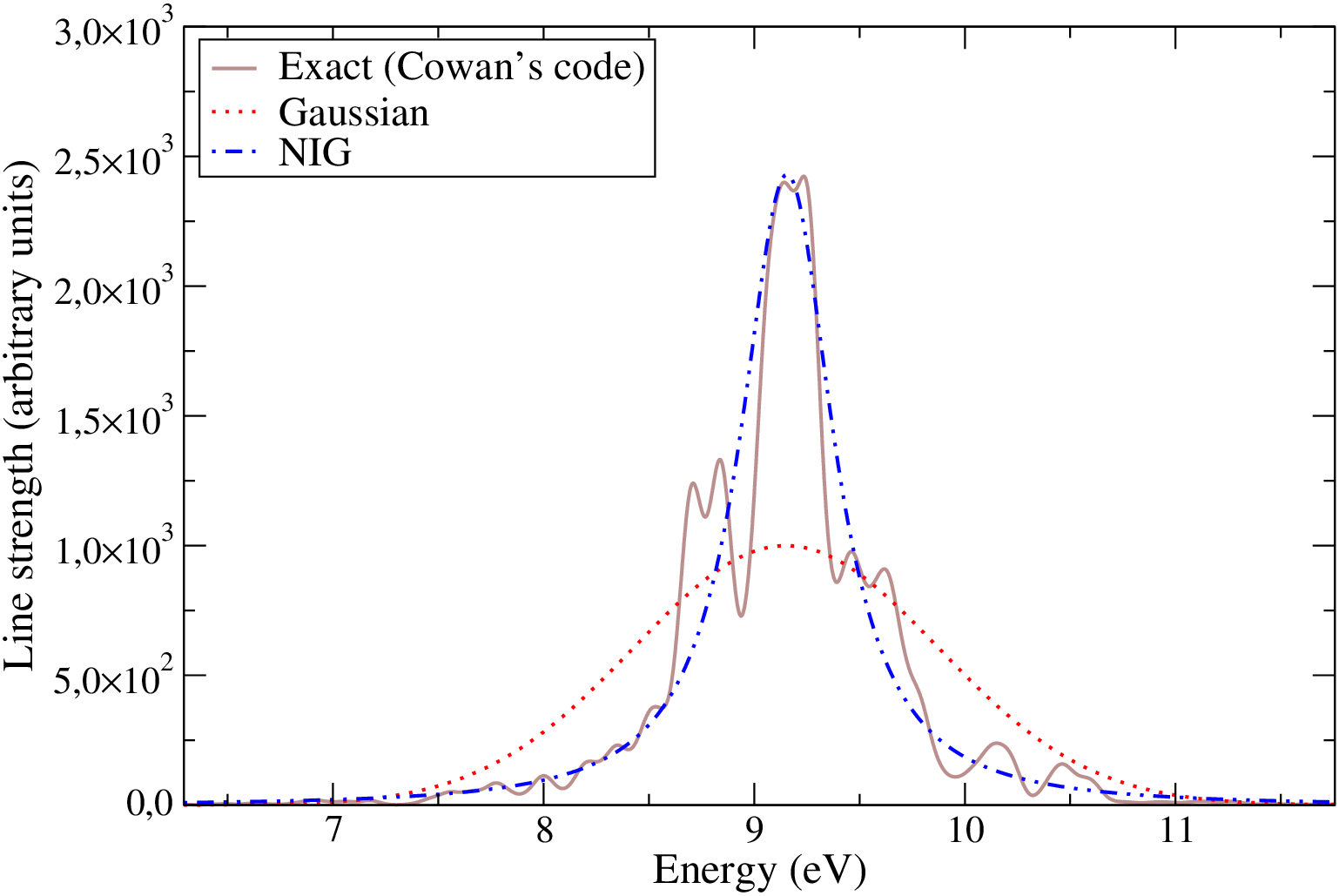}
  \caption{Different modelings of transition arrays $3d^34p\rightarrow 3d^34d$ in Co V.}\label{ggnigb}
\end{center}
\end{figure}

The $n^{th}$-order strength-weighted centered moment of the distribution of lines in a transition array connecting two configurations $C$ and $C'$ reads

\begin{equation}
\mu_n^c=\frac{\sum_{d,u}S_{du}\left(E_u-E_d\right)^n}{\sum_{d,u}S_{du}},
\end{equation}

\noindent where $S_{du}$ is the strength of the line $d\rightarrow u$, $d$ being a level of $C$ and $u$ a level of $C'$, with energies $E_d$ and $E_u$ respectively. One of the main assumptions of the UTA (Unresolved Transition Array) formalism \cite{Bauche-Arnoult79-82-85} is the use of a Gaussian function for representing a transition array of lines. We found that the Normal Inverse Gaussian (NIG) provides a better description of the profile (especially in the wings, see Fig. \ref{wingsa}, \ref{wingsb}, \ref{ggniga} and \ref{ggnigb}) \cite{Pain09}. It reads, as a function of the photon energy $E$:

\begin{equation}
\mathcal{A}(E)=\frac{\delta\alpha e^{\delta\sqrt{\alpha^2-\tau^2}+\tau(E-\chi)}}{\pi\sqrt{\delta^2+(E-\chi)^2}}K_1\left(\alpha\sqrt{\delta^2+(E-\chi)^2}\right),
\end{equation}

\noindent where $K_1(z)$ is a modified Bessel function of the second kind (solution of $z^2y''+zy'-(z^2+1)y=0$) and the parameters $\alpha$, $\tau$, $\delta$ and $\chi$ are determined from the knowledge of the moments. The latests can be expressed in terms of sums involving products of radial integrals. If the transition array is characterized by $q$ different Slater integrals $R$ and $r$ different spin-orbit integrals $\zeta$, the maximal number of terms of the form $\underbrace{R\cdots R}_{m\;\mathrm{terms}}\underbrace{\zeta\cdots\zeta}_{p\;\mathrm{terms}}$ is

\begin{equation}
N_{\mathrm{max}}(n,q,r)=\underbrace{\sum_{m=0}^n\sum_{p=0}^n}_{m+p=n}\sum_{i=0}^m\sum_{j=0}^pS_q(i)\times S_r(j)\;\;\;\;\mathrm{where}\;\;\;\;S_t(v)=\bin{v+t-2}{t-2},
\end{equation}

\noindent which implies 

\begin{equation}
N_{\mathrm{max}}(n,q,r)=\frac{(n+q+r-1)!}{n!(q+r-1)!}.
\end{equation}

\noindent Table \ref{tab:a} gives the maximal number of terms of the first ten moments of $\ell^ {N+1}\rightarrow\ell^ N\ell'$ (for which $q=4$ and $r=3$). We have withdrawn the number of terms of the kind $R\cdots R\zeta$ (equal to $n(n+1)(n+2)/2$) which contribution is zero, yielding $\tilde{N}_{\mathrm{max}}(n,4,3)=N_{\mathrm{max}}(n,4,3)-n(n+1)(n+2)/2$. 

\begin{table}[h]
\begin{center}
\tabcolsep7pt
\begin{tabular}{lcccccccccc}\hline
$\mathbf{order}$ $n$ & 1 & 2 & 3 & 4 & 5 & 6 & 7 & 8 & 9 & 10\\\hline
$\tilde{N}_{\mathrm{max}}(n,4,3)$ & 4 & 16 & 54 & 150 & 357 & 756 & 1464 & 2643 & 4510 & 7348\\\hline
\end{tabular}
\caption{Maximum number of terms in the ten first moments of $\ell^ {N+1}\rightarrow\ell^ N\ell'$.}\label{tab:a}
\end{center}
\end{table}

\vspace{5mm}

\vspace{5mm}

\noindent The asymmetry and sharpness of a transition array are characterized repectively by $\alpha_3$ (skewness) and $\alpha_4$ (kurtosis), the reduced-centered moment $\alpha_n$ being defined as

\begin{equation}
\alpha_n=\frac{\mu_n^c}{v_w^{n/2}},
\end{equation}

\noindent where $\mu_n^c$ represents the $n^{th}-$order moment of the line energies weighted by the strengths and $v_{w}=\mu_2^c$ is the variance. The kurtosis can be estimated considering the ideal joint distribution of line energies $\epsilon$ and amplitudes $a$:

\begin{equation}
\mathcal{D}(\epsilon,a)=\frac{L}{\sqrt{2\pi v}}\exp\left[-\frac{\epsilon^2}{2v}\right]\frac{\lambda}{2}\exp\left[-\lambda~|a|\right]\;\;\;\;\mathrm{with}\;\;\;\;\frac{2}{\lambda^2}=\gamma\exp\left[-\eta~|\epsilon|\right],
\end{equation}

\noindent the correlation law between line energies and amplitudes, $v$ being the unweighted variance (sum of the variances of the levels energies of initial and final configurations). Parameters $\gamma$ and $\eta$ are determined by requiring the conservation of total strength and weighted variance $v_{w}$ of the line energies. One finds

\begin{equation}
\alpha_4=\frac{1}{\omega^2}\left[-2+(5+X^2)\omega\right],
\end{equation}

\noindent where $X$ is root of equation

\begin{equation}
\left(1+X^2-\omega\right)\exp\left[\frac{X^2}{2}\right]\mathrm{erfc}\left[\frac{X}{\sqrt{2}}\right]-\sqrt{\frac{2}{\pi}}X=0,
\end{equation}

\noindent with $\omega=v_{w}/v$. For instance, for transition array $3p^33d^5\rightarrow 3p^23d^6$, we obtain $\alpha_4$=4.219 and the exact value is 4.215.

\vspace{5mm}

\subsection{Finite-temperature effects}

In the statistical approach of configurations, the average quantities are usually derived in the high-temperature limit, i.e. using the degeneracy of levels as weight factors. The SWAP approximation (Statistical Weight APproximation) allows one to derive compact formulas using Racah algebra:  UTA moments (mean energy, variance) for absorption or emission rates, collisional rates, etc. We proposed corrections to the SWAP model taking into account the effects of a Boltzmann factor on the population of levels in order to extend the DCA (Detailed Configuration Accounting) model as far as possible to lower temperatures. Racah algebra cannot be used due to the exponential factors which implies that approximations are required. Our approach consists in expressing a configuration average with respect to the high-temperature limit (SWAP), with a correction which contains only average exponential terms. The partition function of configuration $C$ can be put in the form (setting $\beta=1/(k_BT)$):

\begin{equation}
\mathcal{Z}_C[T]=\left(\sum_dg_d\right)\left(\frac{\sum_dg_de^{-\beta E_d}}{\sum_dg_d}\right)=\mathcal{Z}_C[\infty].\langle e^{-\beta E_d}\rangle_{g_d},
\end{equation}

\noindent where $\mathcal{Z}_C[\infty]$ represents the SWAP value and $\langle e^{-\beta E_d}\rangle_{g_d}$ the correction ($d$ is a level of $C$ and $g_d$ its degeneracy). In the same way, the rate between configurations $C$ and $C'$ can be put in the form

\begin{equation}
\mathcal{R}_{C\rightarrow C'}[T]=\frac{\sum_dg_d\mathcal{R}_{d\rightarrow u}e^{-\beta E_d}}{\sum_dg_de^{-\beta E_d}}=\mathcal{R}_{C\rightarrow C'}[\infty].\left(\frac{\langle e^{-\beta E_d}\rangle_{g_d\mathcal{R}_{d\rightarrow u}}}{\langle e^{-\beta E_d}\rangle_{g_d}}\right),
\end{equation}

\noindent where $\mathcal{R}_{d\rightarrow u}$ is the rate of a processus connecting level $d$ of $C$ and level $u$ of $C'$ and 

\begin{equation}
\langle e^{-\beta E_k}\rangle_{w_k}=\left(\sum_kw_ke^{-\beta E_k}\right)\left/\left(\sum_kw_k\right)\right., 
\end{equation}

\noindent $w_k$ being arbitrary weights (degeneracy, rate, strength, etc). For instance, one has (see Fig. \ref{rate}, left):

\begin{equation}
\frac{\mathcal{B}_{C\rightarrow C'}[T]}{\mathcal{B}_{C\rightarrow C'}[\infty]}=\frac{g_C\left(\sum_{d,u}S_{du}e^{-\beta E_d}\right)}{\left(\sum_{d,u}S_{du}\right)\left(\sum_{d\in C}g_de^{-\beta E_d}\right)},
\end{equation}

\noindent where $\mathcal{B}$ is Einstein absorption rate and $S_{du}$ the line strength. Jensen's inequality for the convex exponential function yields $\langle e^{-\beta E_k}\rangle\geq e^{-\beta\langle E_k\rangle}$ so that $\mathcal{Z}_C[T]=\mathcal{Z}_C[\infty].\langle e^{-\beta E_d}\rangle_{g_d}\geq g_c.e^{-\beta E_C}$ and therefore usual approximation always underestimates the partition function. Using the generalized $J$-file sum rule (see Appendix A) for $\ell^{N+1}\ell'^{N'}$, we get

\begin{equation}
\sum_{d,u}S_{du}e^{-\beta E_d}=I_{\ell\ell'}^2\sum_{d\in C}g_de^{-\beta E_d}\left(\frac{(N+1)\ell_>}{2\ell+1}+C\left[G^1\left(\ell\ell'\right);d\right]\right),
\end{equation}

\noindent $C\left[G^1\left(\ell\ell'\right);d\right]$ being the coefficient of $G^1$ Slater integral in the energy of level $d$ of configuration $\ell^{N+1}\ell'^{N'}$, $\ell_>=\max\left(\ell,\ell'\right)$ and $I_{\ell\ell'}$ the radial dipolar integral. We note that if $N'=0$ (transition array $\ell^{N+1}\rightarrow\ell^N\ell'$), there is no $G^1$ integral in configuration $C$ and then

\begin{equation}
\sum_{d,u}S_{du}e^{-\beta E_d}=\frac{1}{g_C}\left(\sum_{d,u}S_{du}\right)\left(\sum_{d\in C}g_de^{-\beta E_d}\right)
\end{equation}

\noindent so that $\mathcal{B}_{C\rightarrow C'}[T]/\mathcal{B}_{C\rightarrow C'}[\infty]=1$ for $\lambda^{\nu}\ell^{N+1}\rightarrow\lambda^{\nu}\ell^N\ell'$ or $\lambda^{\nu}\ell^{4\ell+2}\ell'^ {N'}\rightarrow\lambda^{\nu}\ell^{4\ell+1}\ell'^{N'+1}$ and $<1$ otherwise. The SWAP value gives exact results for important transition arrays (those involving ground states) and overestimates the photo-absorption rate in other cases. 

$\bullet$ In order to evaluate the required corrections, it is necessary to find general approximations for the average exponential terms, regardless of the weight factors $w_k$. We first developed the so-called ``distribution method'' \cite{Gilleron11}, which consists in assuming that level energies vary continuously and  introducing a weighting distribution $D_w[x]$. For instance, with a square distribution $D_w[x]=\frac{1}{2\sqrt{3}\sigma}\theta\left[\sqrt{3}-\frac{|x|}{\sigma}\right]$, one finds

\begin{equation}
\langle e^{-\beta\epsilon_k}\rangle\approx\int_{-\infty}^{\infty}dxD_w[x]e^{-\beta x}=\sinh[\sqrt{3}\beta\sigma]/(\sqrt{3}\beta\sigma).
\end{equation}

$\bullet$ In addition, we tested the second-order Taylor-series expansion \cite{Gilleron11}:

\begin{equation}
\langle e^{-\beta E_k}\rangle\approx e^{-\beta\langle E_k\rangle}\left(1+\frac{\beta^2}{2}\langle\left(E_k-\langle E_k\rangle\right)^2\rangle\right)
\end{equation}

\noindent which yields

\begin{equation}
\frac{\mathcal{R}_{C\rightarrow C'}[T]}{\mathcal{R}_{C\rightarrow C'}[\infty]}\approx e^{-\beta\left(X_{CC'}-E_C\right)}.\left(\frac{2+\beta^2Y_{CC'}}{2+\beta^2\sigma_{C'}^2}\right)
\end{equation}

\noindent where $\sigma_{C'}^2$ is the variance of configuration $C'$ and

\begin{equation}
X_{CC'}=\frac{\sum_{d,u}g_d\mathcal{R}_{d\rightarrow u}E_d}{\sum_{d,u}g_d\mathcal{R}_{d\rightarrow u}}\;\;\;\;\mathrm{and}\;\;\;\;Y_{CC'}=\frac{\sum_{d,u}g_d\mathcal{R}_{d\rightarrow u}\left(E_d-X_{CC'}\right)^2}{\sum_{d,u}g_d\mathcal{R}_{d\rightarrow u}}.
\end{equation}

\noindent Corrections involve the first and second moments of the strength-weighted level energies. For absorption, moments of the ``receptive zone'' of configuration C are:

\begin{equation}
m_{1A}=\sum_{d,u}S_{du}E_d\left/\sum_{d,u}S_{du}\right.\;\;\;\mathrm{and}\;\;\; m_{2A}=\sum_{d,u}S_{du}E_d^2\left/\sum_{d,u}S_{du}\right.
\end{equation}

\noindent with 

\begin{equation}
\sigma_A^2=m_{2A}-\left(m_{1A}\right)^2
\end{equation}

\noindent and for emission, moments of the ``emissive zone'' of configuration $C'$ are:

\begin{equation}
m_{1E}=\sum_{d,u}S_{du}E_u\left/\sum_{d,u}S_{du}\right.\;\;\;\mathrm{and}\;\;\; m_{2E}=\sum_{d,u}S_{du}E_u^2\left/\sum_{d,u}S_{du}\right.
\end{equation}

\noindent with

\begin{equation}
\sigma_E^2=m_{2E}-\left(m_{1E}\right)^2.
\end{equation}

\noindent Expressions for $m_{1A}$ (which corresponds to $X_{CC'}$), $m_{2A}$ (which corresponds to $Y_{CC'}$), $m_{1E}$ and $m_{2E}$ are easily deduced from UTA formulas by selecting terms with products of Slater integrals belonging to the same configuration. The variance of the emissive zone for the general transition array $\ell_1^{N_1}\ell_2^{N_2}\rightarrow\ell_1^{N_1+1}\ell_2^{N_2-1}$ is given in Appendix B.

\vspace{5mm}

\begin{figure}[h]
\begin{center}
\includegraphics[width=202pt]{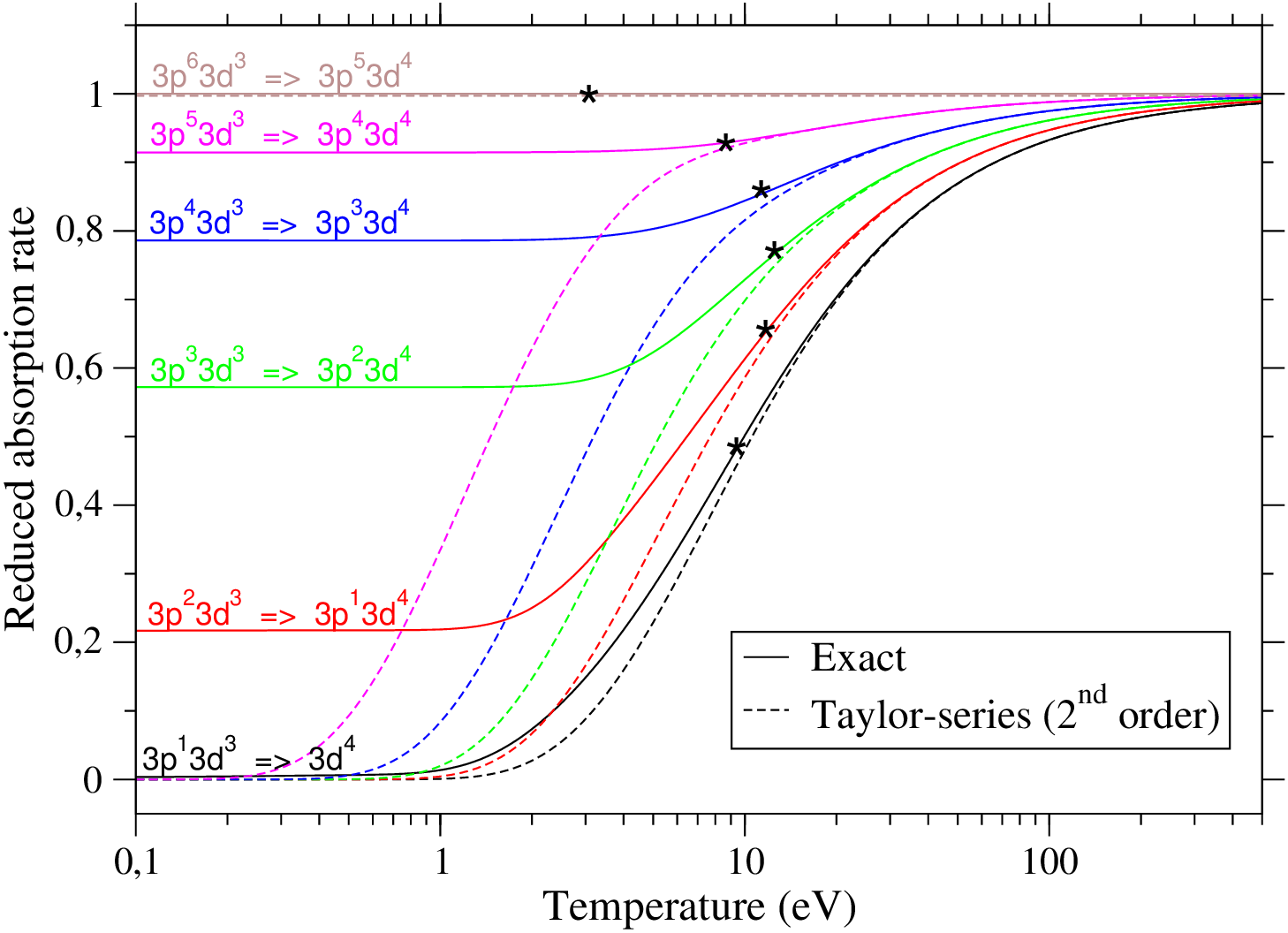}
\caption{Reduced absorption rate as a function of temperature; comparison between exact values (Cowan's code) and second-order Taylor-series approach for $3p^N3d^3\rightarrow 3p^{N-1}3d^4$ with $N$=6 to 2.}\label{ratea}
\end{center}
\end{figure}

\vspace{5mm}

\begin{figure}[h]
\begin{center}
\includegraphics[width=205pt]{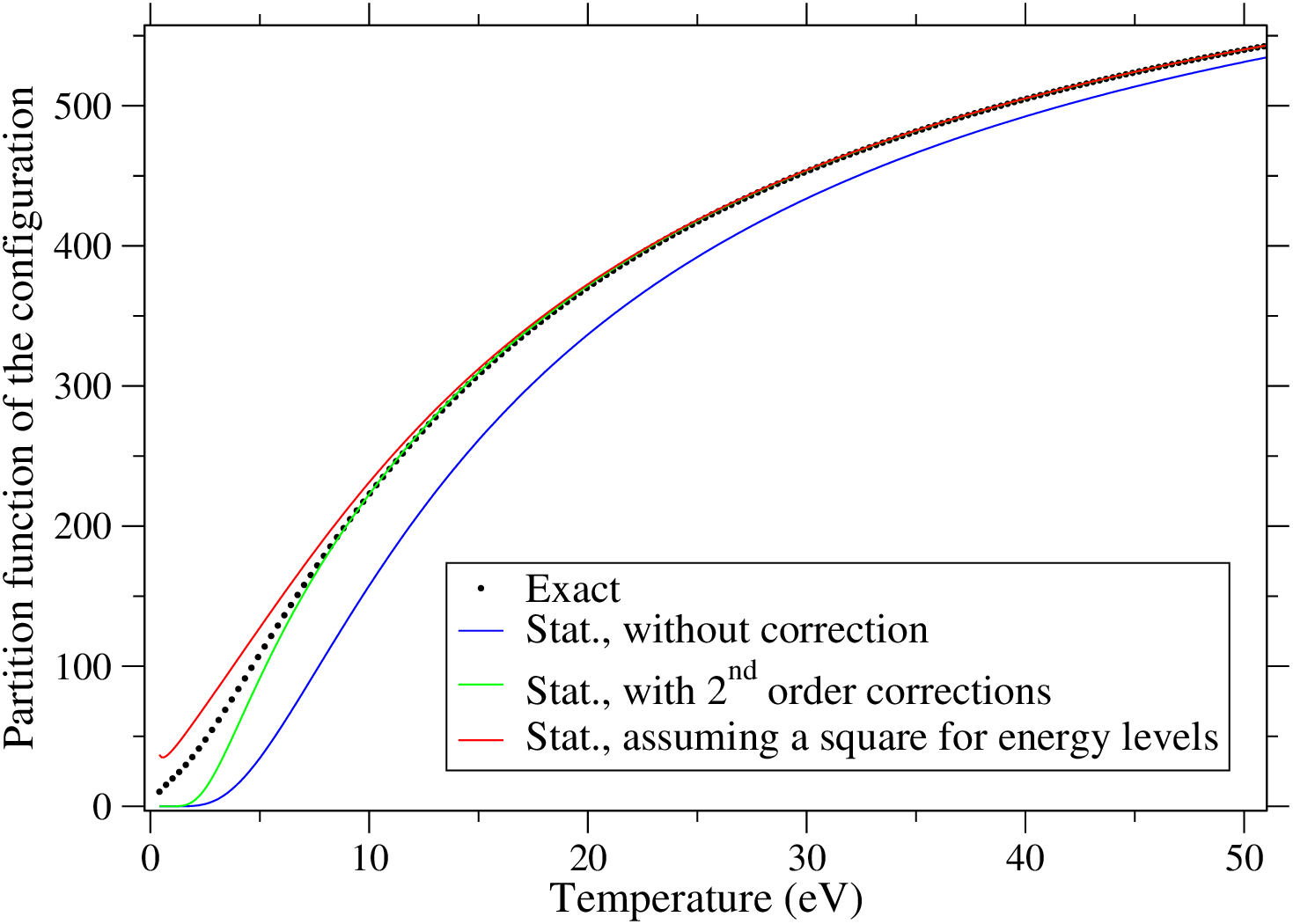}
\caption{Partition fonction as a function of temperature for configuration [Mg]3p$^1$3d$^3$; comparison between exact values (Cowan's code) and different methods proposed here.}\label{rateb}
\end{center}
\end{figure}

\noindent We compared both approaches \cite{Gilleron11} for absorption rate (see Fig. \ref{ratea}) and partition function (see Fig. \ref{rateb}) and the corrections are fairly accurate down to temperatures corresponding to $k_BT\approx\sigma$, i.e. of the order of 10 eV (those conditions for lower configurations are marked as stars in Fig. \ref{ratea}).

\vspace{5mm}

\begin{figure}[h]
\begin{center}
  \includegraphics[width=205pt]{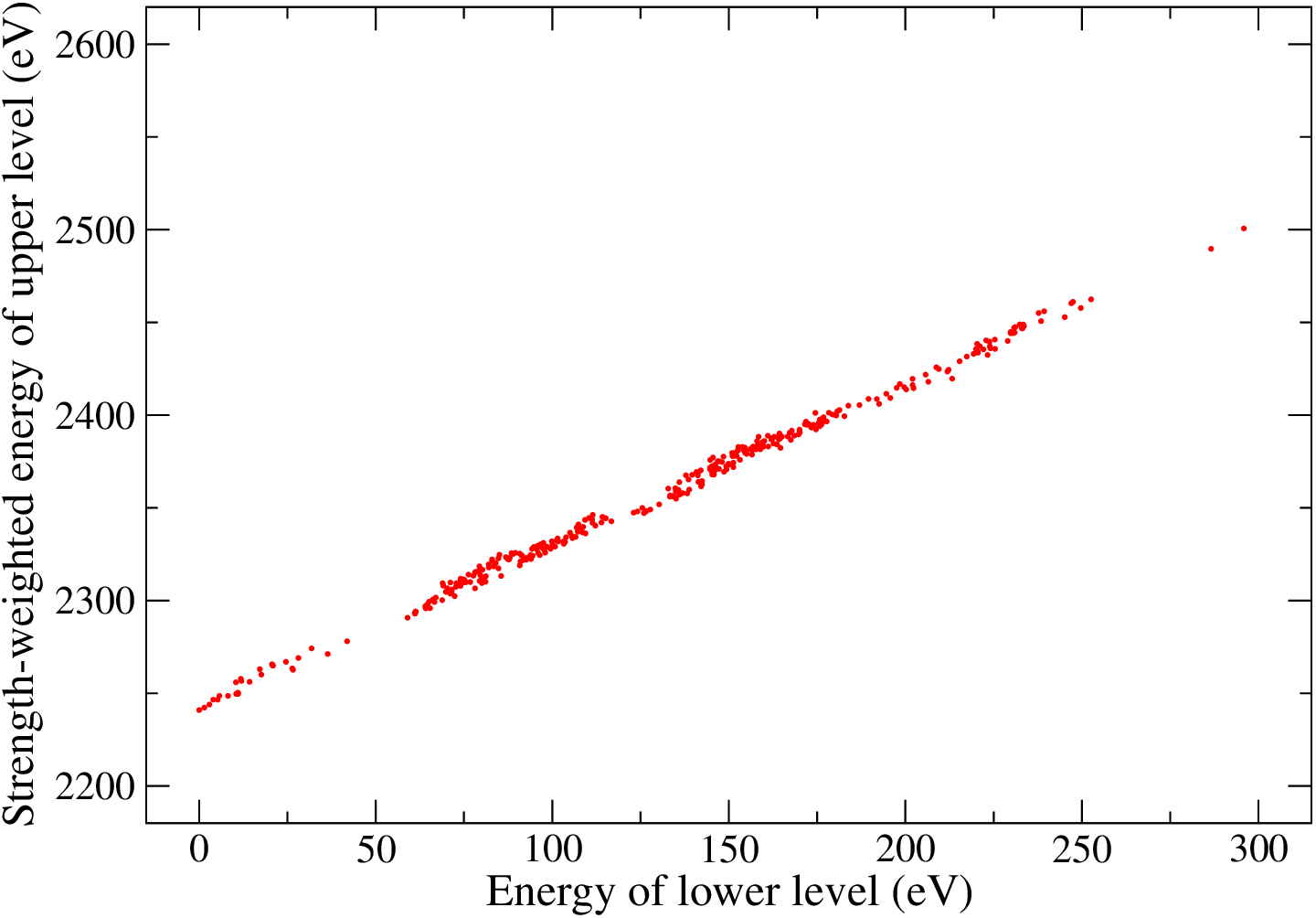}
\caption{Strength-weighted energy of upper level as a function of energy of lower level for $3d^6 4f\rightarrow 3d^5 4f^2$ in W XXXII.}\label{osula}
\end{center}
\end{figure}

\vspace{5mm}

\begin{figure}[h]
\begin{center}
  \includegraphics[width=190pt]{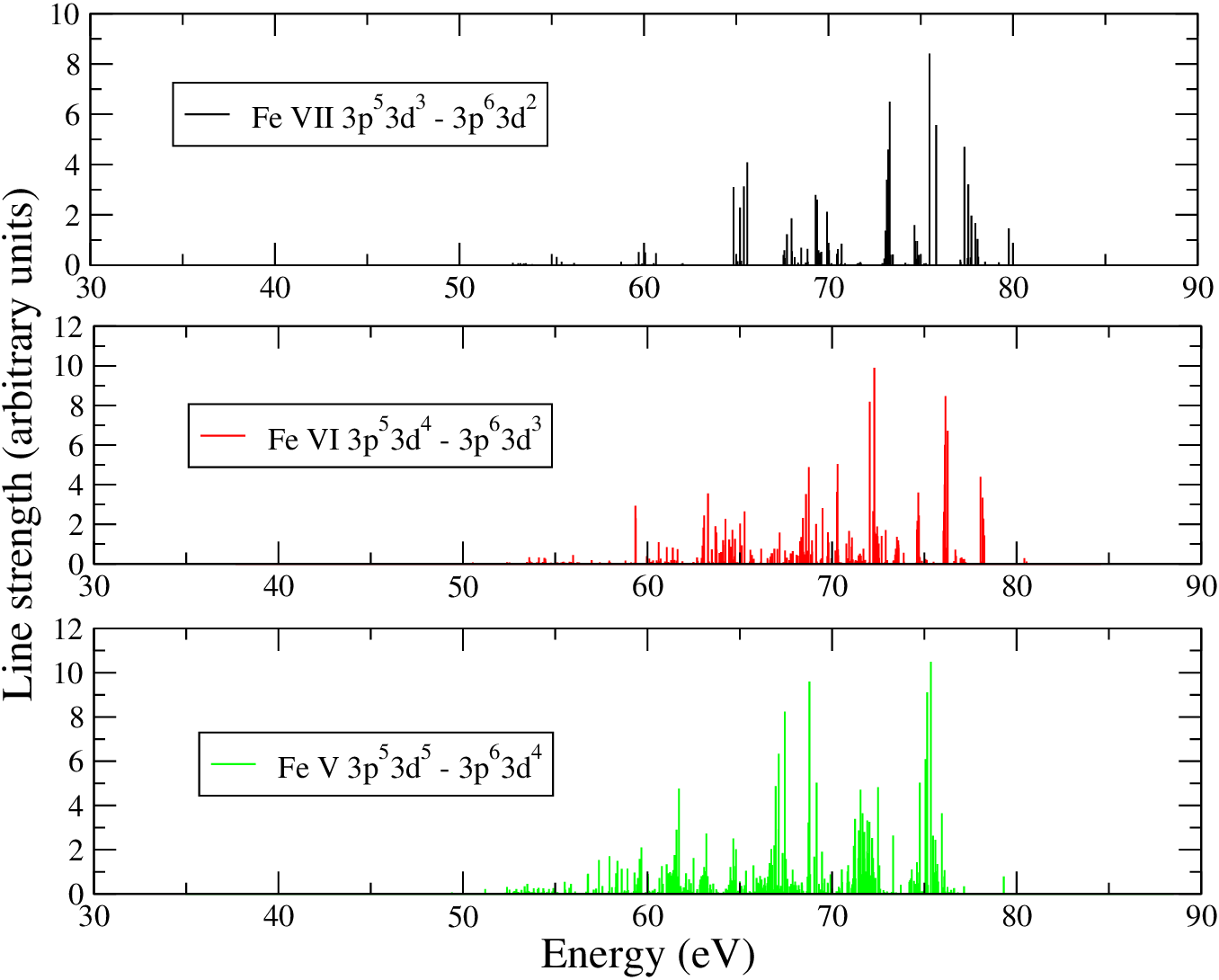}
\caption{The value of the $G^1(3p,3d)$ integral is high and increases as $N$ decreases.}\label{osulb}
\end{center}
\end{figure}

\clearpage

\section{GENERAL PROPERTIES OF THE DISTRIBUTION OF LEVELS AND LINES}

\subsection{Propensity rule}

The propensity rule belongs to a class of tendencies that are well understood by selection rules \cite{Kyniene04,Karazija14} and correlation laws. A line connects preferentially a low (high) energy level of the initial configuration to a low (high) energy level of the final configuration. Such a correlation is due to the selection rules and to the fact that the energies of the levels of a configuration follow a preferential order with respect to the quantum numbers (\emph{cf.} Hund's rule for $\ell^N$ or $\ell^Ns$ configurations, which states that the levels of highest spin have the lowest energy). The selection rule for E1 lines thus imposes a correlation between line energies and their amplitudes \cite{Gilleron07,Bauche91}. Such a correlation may be expressed as   

\begin{equation}\label{cor}
\frac{\int_{-\infty}^{+\infty}dE_dE_d\int_{-\infty}^{+\infty}D\left(E_d,E_u,a\right)a^2da}{\int_{-\infty}^{+\infty}dE_d\int_{-\infty}^{+\infty}D\left(E_d,E_u,a\right)a^2da}\approx K . E_u
\end{equation}

\noindent where $D\left(E_d,E_u,a\right)$ represents the number of lines having an amplitude belonging to $\left[a,a+da\right]$ and connecting an energy level in $\left[E_d,E_d+dE_d\right]$ to an energy level in $\left[E_u,E_u+dE_u\right]$. Equation (\ref{cor}) illustrates the existence of a correlation coefficient $K$, which is equal to

\begin{equation}
K=\frac{\langle E_d.E_u\rangle}{\langle E_u^2\rangle}\;\;\;\;\mathrm{with}\;\;\;\;\langle X\rangle=\frac{\sum_{d,u}XS_{du}}{\sum_{d,u}S_{du}},
\end{equation}

\noindent the latest quantity being the strength-weighted average value of quantity $X$. The consequence of such a correlation is that the strongest lines are mostly located around the center of gravity of the transition array (see Fig. \ref{osula}).

\section{About the role of the $G^1$ exchange Slater integral in atomic spectroscopy}\label{sec5}

$\bullet$ In practice the propensity rule is obviously not always satisfied, and one can observe a concentration of the oscillator strength towards the high-energy side of the transition array \cite{OSullivan99}, see Fig. \ref{osulb}, which occurs as well for some complex Auger spectra. For configurations $\ell^N\ell'^{N'+1}$ with two open subshells having the same principal quantum number, the Coulomb exchange interaction energy mainly determines the energy level spectrum. This interaction forms the upper and lower groups of levels with very different abilities to participate in transitions. Due to the relation between the energy of a level and the transition amplitude from this level, the lines mainly from the upper group of levels manifest themselves in the radiative or Auger spectra. The resulting asymmetrical shape of the transition array \cite{Pain13} is linked with the existence of the emissive zones. In general, this goes hand in hand with a dominant exchange Slater integral $G^1$ with a positive coefficient, which is always the case in $\ell^{N+1}\rightarrow \ell^N\ell'$ arrays that are therefore always asymmetrical. The coefficients of $G^1$ in $\ell^N\ell'$ are given by the generalized $J$-file sum rule mentioned above:

\begin{equation}
\mathcal{C}\left[G^1\left(\ell\ell'\right);\alpha J\right]=-\frac{N\ell_>}{(2\ell+1)}+\frac{1}{(2J+1)I_{\ell\ell'}^2}\sum_{\alpha'J'}S_{\alpha J\rightarrow \alpha'J'}.
\end{equation}

\noindent The coefficients of $G^1$ in a configuration $\ell^{4\ell+1}\ell'^{N'}$ or $\ell^N\ell'^{4N'+2}$ are constant and one has

\begin{equation}
\mathcal{C}\left[G^1\left(\ell^{4\ell+2}\ell'^ {N'}\right);\alpha J\right]=-\frac{N'\ell_>}{2\ell'+1}\;\;\; \mathrm{and}\;\;\;\mathcal{C}\left[G^1\left(\ell^N\ell'^ {4\ell'+2}\right);\alpha J\right]=-\frac{N\ell_>}{2\ell+1}.
\end{equation}

\noindent The contribution of $G^1\left(\ell\ell'\right)$ to the variance of the configuration $\ell^N\ell'$ is

\begin{equation}
\sigma^2=\frac{N(4\ell+2-N)}{[\ell,\ell'](4\ell+1)}\;\ell_>^2\;\left(\frac{1}{3}-\frac{1}{4[\ell,\ell']}\right)[G^1(\ell,\ell')]^2
\end{equation}

\noindent where $[x]=2x+1$ and to the third-order centered moment

\begin{eqnarray}
\mu_3^c&=&\frac{N(4\ell+2-N)}{8\ell[\ell,\ell'](4\ell+1)}\;\ell_>^3\;\left[\frac{4}{9}(N-1)-(4\ell+1-N)\neufj{\ell}{\ell'}{1}{\ell'}{1}{\ell}{1}{\ell}{\ell'}\right.\nonumber\\
& &+\left.\frac{(2\ell+1-N)}{[\ell,\ell']}\left(2-\frac{1}{[\ell,\ell']}\right)\vphantom{\neufj{\ell}{\ell'}{1}{\ell'}{1}{\ell}{1}{\ell}{\ell'}}\right][G^1(\ell,\ell')]^3.
\end{eqnarray}

$\bullet$ While many measurements have been made of lifetimes in atoms and ions and of branching fractions for transitions in the visible region for neutral atoms, very few information is currently available concerning branching ratios for charged ions or for transitions in the UV region. The reasons for this lack of data involve special difficulties associated with the relative calibration of detection systems. However, semiempirical methods exist for treating divalent systems that utilize singlet-triplet intermediate-coupling amplitudes obtained from spectroscopic energy levels to predict relative transition probability rates. The technique consists in determining singlet-triplet mixing angles from measured energy-level data in two-valence electron systems, and using these mixing angles to specify E1 and M1 oscillator strengths and magnetic Land\'e factors from data obtained in single-valence electron systems. While this requires that both the upper and lower configuration state vectors be dominated by a single configuration, the formulation can effectively characterize the effects of spin-orbit, spin-other-orbit, and indirect configuration interaction effects that may induce differences between the singlet and triplet radial wave functions. The $nsn'\ell$ configuration consists of four levels which, in the limit of pure LS coupling, can be denoted by the standard spectroscopic symbols $^3L_{\ell'}$, $^1L_{\ell'}$, $^3L_{\ell}$ and $^1L_{\ell}$. In intermediate coupling, the physical levels can be described by the wave functions

\begin{equation}
|^3L_{\ell}'\rangle=|^3L_{\ell}\rangle\cos(\theta)-|^1L_{\ell}\rangle\sin(\theta)\;\;\;\;\mathrm{and}\;\;\;\;|^1L_{\ell}'\rangle=|^3L_{\ell}\rangle\sin(\theta)+|^1L_{\ell}\rangle\cos(\theta),
\end{equation}

\noindent where $\theta$ is the singlet-triplet mixing angle. The mixing can be expressed in terms of the Slater exchange energy parameter $G_1$ ($G^1$ divides by a specific tabulated coefficient \cite{Condon35}) and the diagonal and off-diagonal magnetic energy parameters $\delta_1$ and $\delta_2$ as

\begin{equation}
\mathrm{cotan}(2\theta)=\frac{2G_1+\delta_1/2}{\sqrt{\ell(\ell+1)}~\delta_2}.
\end{equation}

\noindent In the simplest formulation \cite{Curtis89,Curtis00}, the diagonal and off-diagonal magnetic parameters are both set equal to the standard spin-orbit energy yielding $\delta_1=\delta_2=\zeta$ but this formulation can be extended to include the spin-other-orbit interaction energy \cite{Wolfe32}. Such a formalism provides a predictive systematization of lifetime and energy level data that is simple to use, contains internal checks of its validity, and permits predictions of lifetimes, branching fractions, and transition probabilities.

\subsection{Estimating the number of lines in a transition array}

In some situations, the distribution of the values of $M$ (projection of angular momentum $J$) exhibits a ``plateau'', which is reproduced neither by a Gaussian nor by Gram-Charlier (GC) distribution. The Generalized Gaussian (GG):

\begin{equation}\label{nli}
P(M)=\frac{g_C}{2\lambda\sigma\Gamma\left(1+1/n\right)}e^{-|u/\lambda|^n}\;\;\; \mathrm{with}\;\;\; u=M/\sigma\;\;\; \mathrm{and}\;\;\; \lambda=\sqrt{\frac{\Gamma\left(1/n\right)}{\Gamma\left(3/n\right)}},
\end{equation}

\noindent where $\Gamma$ represents the usual Gamma function and $\sigma$ the variance of $P(M)$ \cite{Gilleron09}, provides a better description of $P(M)$ \cite{Gilleron09} and we developed a new approximate formula for the number of electric-dipole (E1) lines in a transition array  with a wider range of applicability \cite{Gilleron09}:

\begin{equation} 
N_{\mathrm{lines}}=\frac{2^{1/n}(n-1)g_Cg_{C'}}{64\lambda^5\sigma^5\Gamma\left(1+1/n\right)^2}\left[12\lambda^2\sigma^2\Gamma\left(1-\frac{1}{n}\right)+2^{2/n}n(1-2n)\Gamma\left(2-\frac{3}{n}\right)\right].
\end{equation}

\noindent For instance, for $d^7\rightarrow d^6f$, the exact number of lines is 2825 and the values obtained with a Gram-Charlier and a GG modelings of $P(M)$ are respectively 2859 and 2845; for $d^4i\rightarrow d^3ip$, the exact number of lines is 193735 and the values obtained with a Gram-Charlier and a GG modelings of $P(M)$ are respectively 187549 and 198690. Estimating the number of lines is useful for hybrid atomic-structure codes \cite{Pain15a,Pain15b}, in order to decide whether a transition array should (could) be detailed or not.

\section{CONCLUSION}

Detailed calculations show that the Gaussian is not always the most relevant distribution for modeling transition arrays of E1 lines and that the Normal Inverse Gaussian seems more suited in many cases. Corrections to extend the validity range of statistical methods (UTA, SOSA) towards low or moderate temperatures were proposed. Regularities and symmetry properties, in addition to their fundamental interest \cite{Bauche97}, can be used as constraints for approximate models. Efficient techniques were developed for calculating the distribution of quantum numbers in a configuration and the number of lines in a transition array. We believe that group theory will certainly help making further significant progress \cite{Weyl50,Wybourne70,Judd96} in the field.

\section{ACKNOWLEDGMENTS}

We would like to thank Jacques Bauche and Claire Bauche-Arnoult for their work, which is for us an invaluable source of inspiration.

\section{Appendix A: J-file sum rules for E1 and E2 lines}

The $J$-file sum rule for E1 lines reads \cite{Bauche83}:

\begin{equation}
S_{E_1}\left[\left(\ell^N\ell'^{N'+1}\right)\alpha J\rightarrow\ell^{N+1}\ell'^{N'}\right]=(2J+1)\left(\frac{(N'+1)}{2\ell'+1}\langle\ell||C^{(1)}||\ell'\rangle^2+C\left[G^1\left(\ell\ell'\right);\alpha J\right]\right)I_{\ell\ell'}^2
\end{equation}

\noindent where $I_{\ell\ell'}=\int_0^{\infty}R_{n\ell}(r)rR_{n'\ell'}(r)dr$ and $\langle\ell||C^{(1)}||\ell'\rangle^2=\ell_>=\max\left(\ell,\ell'\right)$. The coefficient $C\left[G^1\left(\ell\ell'\right);\alpha J\right]$ represents the coefficient of $G^1$ in the energy of level $\alpha J$ of configuration $\ell^N\ell'^{N'+1}$. For E2 lines, one has \cite{Pain12,Bauche15}:

\begin{equation}
S_{E_2}\left[\left(\ell^N\ell'^{N'+1}\right)\alpha J\rightarrow\ell^{N+1}\ell'^{N'}\right]=(2J+1)\left(\frac{(N'+1)}{2\ell'+1}\langle\ell||C^{(2)}||\ell'\rangle^2+C\left[G^2\left(\ell\ell'\right);\alpha J\right]\right)J_{\ell\ell'}^2
\end{equation}

\noindent where $J_{\ell\ell'}=\int_0^{\infty}R_{n\ell}(r)r^2R_{n'\ell'}(r)dr$, $C\left[G^2\left(\ell\ell'\right);\alpha J\right]$ is the coefficient of $G^2$ in the energy of level $\alpha J$ of configuration $\ell^N\ell'^{N'+1}$ and 

\begin{equation}
\left\{\begin{array}{lll}
\mathrm{if} & |\ell-\ell'|=2, & \langle\ell||C^{(2)}||\ell'\rangle^2=\frac{3\ell_>(\ell_>-1)}{2(2\ell_>-1)}\\
\mathrm{if} & \ell=\ell', & \langle\ell||C^{(2)}||\ell'\rangle^2=\frac{\ell(\ell+1)(2\ell+1)}{(2\ell-1)(2\ell+3)}
\end{array}\right.. 
\end{equation}

\section{Appendix B: Variance of the emissive zone of the transition array : $C=C_0~\ell_1^{N_1}\ell_2^{N_2}\rightarrow C'=C_0~\ell_1^{N_1+1}\ell_2^{N_2-1}$}

\noindent The expression of the variance of the emissive zone of the transition array : $C_0~\ell_1^{N_1}\ell_2^{N_2}\rightarrow C_0~\ell_1^{N_1+1}\ell_2^{N_2-1}$ (with spectators $C_0$) can be obtained from reference \cite{Karazija88}, by discarding all radial integrals of the lower configuration in the expression of the variance of the transition array. It can be put in the form $\sigma^2(C-C')=\sigma^2(C)+\delta\sigma^2(C)-(\delta y(C))^2$. Using the same convention as in Ref. \cite{Karazija88}, i.e. 

\begin{equation}
\left\{
\begin{array}{lll}
\ff{k}{\ell}{\ell'}&=&(-1)^{\ell+\ell'+k}\left[\ell,\ell'\right]\troisj{\ell}{k}{\ell}\troisj{\ell'}{k}{\ell'}F^k\left(\ell\ell'\right)\\
\gd{k}{\ell}{\ell'}&=&(-1)^{\ell+\ell'+k}\left[\ell,\ell'\right]\troisj{\ell}{k}{\ell'}^2G^k\left(\ell\ell'\right)
\end{array},\right.
\end{equation}

\noindent we have

\begin{eqnarray}
\sigma^2(C)&=&\sum_i\left\{\frac{N_i(N_i-1)(4\ell_i+2-N_i)(4\ell_i+1-N_i)}{(4\ell_i+2)(4\ell_i+1)4\ell_i(4\ell_i-1)}\right.\nonumber\\
& &\times\sum_{k>0,~k'>0}\left[\frac{2\delta_{k,k'}}{2k+1}-\sixj{\ell_i}{\ell_i}{k}{\ell_i}{\ell_i}{k'}-\frac{1}{(2\ell_i+1)(4\ell_i+1)}\right]\ff{k}{\ell_i}{\ell_i}\ff{k'}{\ell_i}{\ell_i}\nonumber\\
& &\left.+\frac{N_i(4\ell_i+2-N_i)}{4(4\ell_i+1)}\ell_i(\ell_i+1)\zeta_{\ell_i}^2\right\}+\sum_{i<j}\frac{N_i(4\ell_i+2-N_i)N_j(4\ell_j+2-N_j)}{(4\ell_i+2)(4\ell_i+1)(4\ell_j+2)(4\ell_j+1)}\nonumber\\
& &\times\left\{\sum_{k>0,~k'>0}\frac{4\delta_{k,k'}}{2k+1}\ff{k}{\ell_i}{\ell_j}\ff{k'}{\ell_i}{\ell_j}\right.\nonumber\\
& &+\sum_{k,~k'}\left[\frac{4\delta_{k,k'}}{2k+1}-\frac{1}{(2\ell_i+1)(2\ell_j+1)}\right]\gd{k}{\ell_i}{\ell_j}\gd{k'}{\ell_i}{\ell_j}\nonumber\\
& &\left.-4\sum_{k>0,~k'}\sixj{\ell_i}{\ell_i}{k}{\ell_j}{\ell_j}{k'}\ff{k}{\ell_i}{\ell_j}\gd{k'}{\ell_i}{\ell_j}\right\}.
\end{eqnarray}

\noindent The second term $\delta\sigma^2(C)$ reads

\begin{eqnarray}
\delta\sigma^2(C)&=&\frac{N_1(4\ell_2+2-N_2)}{(4\ell_1+1)4\ell_1(4\ell_2+1)4\ell_2}\left\{4(2\ell_1+1-N_1)(2\ell_2+1-N_2)\sigma_e^2(\ell_1^{1},\ell_2^{1})\right.\nonumber\\
& &+[(N_1-1)(4\ell_2+1-N_2)+(4\ell_1+1-N_1)(N_2-1)]M_2(\ell_1,\ell_2)\nonumber\\
& &\left.+[(N_1-1)(N_2-1)+(4\ell_1+1-N_1)(4\ell_2+1-N_2)]M_3(\ell_1,\ell_2)\right\}\nonumber\\
& &-\frac{2N_1(4\ell_2+2-N_2)}{(4\ell_1+1)(4\ell_2+1)}\left[\vphantom{\sum_{p\in C_0}\frac{N_p(4\ell_p+2-N_p)}{(4\ell_p+2)(4\ell_p+1)}M_1(\ell_1\ell_p,\ell_2\ell_p)}\frac{(N_1-1)(4\ell_1+1-N_1)}{4\ell_1(4\ell_1-1)}M_1(\ell_1\ell_1,\ell_1\ell_2)\right.\nonumber\\
& &+\frac{(N_2-1)(4\ell_2+1-N_2)}{4\ell_2(4\ell_2-1)}M_1(\ell_2\ell_2,\ell_1\ell_2)\nonumber\\
& &+\left.\sum_{p\in C_0}\frac{N_p(4\ell_p+2-N_p)}{(4\ell_p+2)(4\ell_p+1)}M_1(\ell_1\ell_p,\ell_2\ell_p)\right]\nonumber\\
& &-\frac{N_1(4\ell_2+2-N_2)}{4(4\ell_1+1)(4\ell_2+1)}[\ell_1(\ell_1+1)+\ell_2(\ell_2+1)-2]\zeta_{\ell_1}\zeta_{\ell_2},
\end{eqnarray}

\noindent where the coefficients $M_i$ are defined as

\begin{eqnarray}
M_1(\ell_1\ell_1,\ell_1\ell_2)&=&\sum_{k>0,~k'>0}\left[\frac{2\delta_{k,k'}}{2k+1}-\sixj{\ell_1}{\ell_1}{k'}{\ell_1}{\ell_1}{k}-\frac{1}{(2\ell_1+1)(4\ell_1+1)}\right]\nonumber\\
& &\times\sixj{\ell_1}{\ell_1}{k'}{\ell_2}{\ell_2}{1}\ff{k}{\ell_1}{\ell_1}\ff{k'}{\ell_1}{\ell_2}\nonumber\\
& &+\sum_{k>0,~k'}\left[2\sixj{k}{k'}{1}{\ell_2}{\ell_1}{\ell_1}^2-\sixj{\ell_2}{\ell_2}{k}{\ell_1}{\ell_1}{1}\sixj{\ell_2}{\ell_2}{k}{\ell_1}{\ell_1}{k'}\right.\nonumber\\
& &+\left.\frac{1}{(2\ell_1+1)(4\ell_1+1)}\left(\frac{2}{3}\delta_{k',1}-\frac{1}{(2\ell_2+1)}\right)\sixj{k}{k'}{1}{\ell_2}{\ell_1}{\ell_1}^2\right]\ff{k}{\ell_1}{\ell_1}\gd{k'}{\ell_1}{\ell_2},\nonumber\\
\end{eqnarray}

\begin{eqnarray}
M_1(\ell_2\ell_2,\ell_1\ell_2)&=&\sum_{k>0,~k'>0}\left[\frac{2\delta_{k,k'}}{2k+1}-\sixj{\ell_2}{\ell_2}{k'}{\ell_2}{\ell_2}{k}-\frac{1}{(2\ell_2+1)(4\ell_2+1)}\right]\nonumber\\
& &\times\sixj{\ell_2}{\ell_2}{k'}{\ell_1}{\ell_1}{1}\ff{k}{\ell_2}{\ell_2}\ff{k'}{\ell_1}{\ell_2}\nonumber\\
& &+\sum_{k>0,~k'}\left[2\sixj{k}{k'}{1}{\ell_1}{\ell_2}{\ell_2}^2-\sixj{\ell_1}{\ell_1}{k}{\ell_2}{\ell_2}{1}\sixj{\ell_1}{\ell_1}{k}{\ell_2}{\ell_2}{k'}\right.\nonumber\\
& &+\left.\frac{1}{(2\ell_2+1)(4\ell_2+1)}\left(\frac{2}{3}\delta_{k',1}-\frac{1}{(2\ell_1+1)}\right)\vphantom{\sixj{k}{k'}{1}{\ell_2}{\ell_1}{\ell_1}^2}\right]\ff{k}{\ell_2}{\ell_2}\gd{k'}{\ell_1}{\ell_2},
\end{eqnarray}

\begin{eqnarray}
M_1(\ell_1\ell_p,\ell_2\ell_p)&=&2\sum_{k>0}\frac{1}{2k+1}\sixj{\ell_1}{\ell_1}{k}{\ell_2}{\ell_2}{1}\ff{k}{\ell_1}{\ell_p}\ff{k'}{\ell_2}{\ell_p}\nonumber\\
& &-\sum_{k>0,~k'}\sixj{\ell_1}{\ell_1}{k}{\ell_2}{\ell_2}{1}\left[\sixj{\ell_1}{\ell_1}{k}{\ell_p}{\ell_p}{k'}\gd{k'}{\ell_1}{\ell_p}\ff{k}{\ell_2}{\ell_p}\right.\nonumber\\
&&+\left.\sixj{\ell_2}{\ell_2}{k}{\ell_p}{\ell_p}{k'}\ff{k}{\ell_1}{\ell_p}\gd{k'}{\ell_2}{\ell_p}\right]\nonumber\\
& &+\sum_{k,k'}\left[2\sixj{\ell_1}{\ell_p}{k}{k'}{1}{\ell_2}^2-\frac{1}{2(2\ell_1+1)(2\ell_2+1)(2\ell_p+1)}\right]\gd{k}{\ell_1}{\ell_p}\gd{k'}{\ell_2}{\ell_p},\nonumber\\
\end{eqnarray}

\begin{equation}
M_2(\ell_1,\ell_2)=\frac{1}{4}\left[-2\sum_{k>0}\sixj{\ell_1}{\ell_1}{k}{\ell_2}{\ell_2}{1}\ff{k}{\ell_1}{\ell_2}+\sum_k\left(\frac{4}{3}\delta_{k,1}-\frac{1}{(2\ell_1+1)(2\ell_2+1)}\right)\gd{k}{\ell_1}{\ell_2}\right]^2,
\end{equation}

\noindent and

\begin{eqnarray}
M_3(\ell_1,\ell_2)&=&-\sum_{k>0,~k'>0}\neufj{\ell_2}{\ell_2}{k}{\ell_2}{1}{\ell_1}{k'}{\ell_1}{\ell_1}\ff{k}{\ell_1}{\ell_2}\ff{k'}{\ell_1}{\ell_2}\nonumber\\
& &+4\sum_{k>0,~k'}\left[\sixj{k}{k'}{1}{\ell_2}{\ell_1}{\ell_1}\sixj{k}{k'}{1}{\ell_1}{\ell_2}{\ell_2}\right.\nonumber\\
& &-\left.\frac{1}{4(2\ell_1+1)(2\ell_2+1)}\sixj{\ell_1}{\ell_1}{k}{\ell_2}{\ell_2}{1}\right]\ff{k}{\ell_1}{\ell_2}\gd{k'}{\ell_1}{\ell_2}\nonumber\\
& &-\sum_{k,~k'}\left[\neufj{\ell_1}{\ell_2}{k}{\ell_2}{1}{\ell_1}{k'}{\ell_1}{\ell_2}-\frac{2\delta_{k',1}}{3(2\ell_1+1)(2\ell_2+1)}\right.\nonumber\\
& &+\left.\frac{1}{4(2\ell_1+1)^2(2\ell_2+1)^2}\vphantom{\neufj{\ell_1}{\ell_2}{k}{\ell_2}{1}{\ell_1}{k'}{\ell_1}{\ell_2}}\right]\gd{k}{\ell_1}{\ell_2}\gd{k'}{\ell_1}{\ell_2}.
\end{eqnarray}

\noindent Finally, the last term $\delta y(C)$, which comes from the shift, reads

\begin{eqnarray}
\delta y(C)&=&\frac{1}{(4\ell_1+1)(4\ell_2+1)}\left[\sum_k\left(\frac{2}{3}\delta_{k,1}-\frac{1}{2(2\ell_1+1)(2\ell_2+1)}\right)N_1(4\ell_2+2-N_2)\gd{k}{\ell_1}{\ell_2}\right.\nonumber\\
& &\left.-\sum_{k>0}\sixj{\ell_1}{\ell_1}{k}{\ell_2}{\ell_2}{1}N_1(4\ell_2+2-N_2)\ff{k}{\ell_1}{\ell_2}\right].
\end{eqnarray}

\end{document}